\documentclass[11pt,a4paper,english,superscriptaddress,aps,preprint]{revtex4}

\usepackage{color}
\usepackage[colorlinks=true, a4paper=true, pdfstartview=FitV,
linkcolor=blue, citecolor=blue, urlcolor=blue]{hyperref}

\usepackage{amsmath}
\usepackage{amssymb}
\usepackage{graphicx}
\usepackage{hhline}
\usepackage{textcomp}
\makeatletter
\usepackage{babel}
\newcommand{\bea}{\begin{eqnarray}}
\newcommand{\eea}{\end{eqnarray}}

\newcommand{\pa}{\partial}

\newcommand{\be}{\begin{equation}}
\newcommand{\ee}{\end{equation}}

\newcommand{\qed}{\nobreak \ifvmode \relax \else
      \ifdim\lastskip<1.5em \hskip-\lastskip
      \hskip1.5em plus0em minus0.5em \fi \nobreak
      \vrule height0.75em width0.5em depth0.25em\fi}
\numberwithin{equation}{section}

\begin{document}
\immediate\write16{<<WARNING: LINEDRAW macros work with emTeX-dvivers
                    and other drivers supporting emTeX \special's
                    (dviscr, dvihplj, dvidot, dvips, dviwin, etc.) >>}

\title{\boldmath Non-uniqueness of the supersymmetric extension of the $O(3)$ $\sigma$-model}
\preprint{KA-TP-09-2017}
\author{Jose M. Queiruga}
\email{jose.queiruga@kit.edu}
\affiliation{Institute for Theoretical Physics, Karlsruhe Institute
of Technology (KIT), 76131 Karlsruhe, Germany}
\affiliation{Institut f\"{u}r Kernphysik, Karlsruhe Institute of Technology, Hermann-von-Helmholtz-Platz 1,
D-76344 Eggenstein-Leopoldshafen, Germany}
 
\author{A. Wereszczynski}
\affiliation{ Institute of Physics, Jagiellonian University, Lojasiewicza 11, Krak\'{o}w, Poland}

\begin{abstract}
We study the supersymmetric extensions of the $O(3)$ $\sigma$-model in $1+1$ and $2+1$ dimensions. We show that it is possible to construct non-equivalent supersymmetric versions of a given model sharing the same bosonic sector and free from higher-derivative terms. 
\end{abstract}

\maketitle


\section{Introduction}
The bosonic nonlinear $O(n)$ $\sigma$-model is probably one of the most studied examples of models where the field space (target space) possesses a nonlinear structure.  
From a physical point of view, one of the main motivations for the investigation of the lower-dimensional nonlinear $\sigma$-models is that they share many similarities with the four-dimensional gauge theories \cite{Polyakov}. Besides, these models are simpler that their four-dimensional counterparts and therefore, they constitute a good laboratory for testing theoretical ideas. 

If one consider the supersymmetric version of these models, a rich mathematical structure arises. For example, the relation between the amount of supersymmetry ($N=1,2,4...$) and the geometrical structure of the target space manifold has been established in \cite{Gaume,Zumino}. It is also well-known that in a number of field theories, the classical solutions can be computed by solving a first-order equation rather than the second-order equation obtained from the variation of the Lagrangian. When this happens, the solutions satisfying the first-order equation saturate a lower-bound for the energy (Bogomolny bound). It has been pointed out  \cite{Witten, Witten1,diVecchia} that this phenomenon has a close relationship with supersymmetry. In a theory with $N=1$ supersymmetry, if a classical topologically nontrivial solution satisfies a first-order equation there exist and extra ($N=2$) supersymmetry. This peculiarity allows in certain cases to determine the exact mass spectrum \cite{Witten1}. Moreover, due to the fact that SUSY relates bosonic and fermionic states, one can obtain, for example fermionic solutions in terms of bosonic ones without solving the corresponding Dirac equation.

Another of the fundamental bridges between supersymmetry and geometry has been discovered in \cite{Witten2}. In these works, the relation between fermionic and bosonic zero-modes (via the Witten index) in the nonlinear $\sigma$-models and topological invariants of the underlying manifolds has been established.

It is the goal of this work to study more general forms of the supersymmetric nonlinear $\sigma$-models with special emphasis on the $O(3)$ model. The classical approaches to the supersymmetric version of these models can be realized in two languages, namely $N=1$ or $N=2$ superspace. In the first case, all the information is encoded in terms of $N=1$ real superfields ($\Phi^k$) (which can be combined eventually in complex superfields). The Lagrangian takes the following form
\be
\mathcal{L}_\sigma^{N=1}=\int d^2\theta g_{ij}(\Phi^k)D^\alpha \Phi^i D_\alpha \Phi^j\label{n1}
\ee
where $g_{ij}(\Phi^k)$ corresponds to the metric on the target-space manifold $\mathcal{M}_T$, in its bosonic restriction. In the $N=2$ language, the Lagrangian (\ref{n1}) is even simpler
\be
\mathcal{L}_\sigma^{N=2}=\int d^2\theta d^2\bar{\theta}K(\Phi^k,\Phi^{k\,\dagger}) \label{n2}
\ee
where $K(\Phi^k,\Phi^{k\,\dagger}) $ is the so-called K\"{a}hler potential (the metric on $\mathcal{M}_T$ can be obtained as the second derivative of the K\"{a}hler potential, $g_{i,\bar{j}}(\phi^k,\phi^{k\,\dagger})=\pa^2_{\Phi^i,\Phi^{j\,\dagger}}K(\Phi^k,\Phi^{k\,\dagger})\vert_{\theta=\bar{\theta}=0}$ ). The superfields $\Phi^k$ are chiral superfields and verify the constraint $\bar{D}_{\dot{\alpha}}\Phi^k=0$. It is well-known that when $\mathcal{M}_T$ is a K\"{a}hler  manifold, the action (\ref{n1}) possesses an extra supersymmetry \cite{Gaume,Zumino}. For an appropriate choice of $K$ one has $\mathcal{L}_\sigma^{N=1}=\mathcal{L}_\sigma^{N=2}$. Or in other words, once one has a  K\"{a}hler  $\sigma$-model (i.e. a $\sigma$-model with a K\"{a}hler target space manifold), the classical formulations (\ref{n1}) and (\ref{n2}) lead irreparably to extra supersymmetry.

The main aim of the current work is to analyze the existence (and properties) of non-equivalent SUSY extensions, which we call bosonic twins \footnote{The term twin-like model was previously used in the literature \cite{twin1,twin2,twin3, twin4} to refer to couples of theories sharing the same topological defect solution with the same energy. Here it is used is a different sense: pairs of supersymmetric theories sharing in the same bosonic sector.}, for a given two-dimensional target-space manifold (for example $O(3)\cong \mathbb{S}^2$). We will see through this work that is is possible to construct supersymmetric versions of the  nonlinear $\sigma$-models with $N=1$ SUSY which do not allow for extra supersymmetry. Further, we will show that in fact, it is possible to generate an infinite family of well-behaved SUSY extensions (in the sense that they do not possess higher-derivative terms) labeled by an arbitrary function. We will also show that, due to the constraints imposed by supersymmetry, and despite of the modification of the fermionic sector, certain solutions of the Dirac equation (fermionic zero-modes) will remain invariant (with respect to (\ref{n1}) and (\ref{n2})). 
 
 \vspace*{0.2cm}
 
 This work is organized as follows. In Sec. 2, we review the SUSY nonlinear $O(3)$-model in terms of real fields and in its $\mathbb{C}P^1$ formulation. In Sec. 3, we introduce the deformation term which allows for the generation of a new fermionic sector for general nonlinear $\sigma$-models. In Sec. 4, we describe the nonlinear $\sigma$-model with potential in $1+1$ dimensions  and determine the fermionic zero-modes from SUSY. In Sec. 5, we determine the full on-shell action with the deformation term and discuss the fermionic zero-modes. In Sec. 6, we describe the pure $O(3)$-model in $2+1$ dimensions (a potential is not allowed) and study fermionic zero-modes and some peculiarities of the quartic fermionic Lagrangian in the presence of the deformation term. In Sec. 7, we describe some properties of the bosonic twin with $N=2$ SUSY. Finally, Sec. 8 is devoted to our summary.
\label{sec:intro}


\section{The $O(3)$ nonlinear $\sigma$-model }

This section is intended for a review of the SUSY $O(3)$ nonlinear $\sigma$-model in two formulations, the $O(3)$ and the $\mathbb{C}P^1$. In the first formulation the model can be written in terms of three real scalar fields $\phi^a$ satisfying one constraint
\be
\sum_{a=1}^3\phi^a \phi^a=1.\label{cons}
\ee
In its non-SUSY version can be written simply as follows 
\be
S=-\frac{1}{2}\int d^2x\pa_\mu \phi^a \pa^\mu \phi^a, \quad \sum_{a=1}^3\phi^a \phi^a=1. \label{o3}
\ee 

The $N=1$ supersymmetric extension of (\ref{o3}) is well-known \cite{Witten}. It only involves three real superfields and the supersymmetric generalization of (\ref{cons}), we have
\be
\mathcal{S}=-\frac{1}{2}\int d^2 xd^2\theta D^\alpha \Phi^a D_\alpha \Phi^a, \quad \sum_{a=1}^3\Phi^a \Phi^a=1\label{so3}
\ee
where $\Phi^a=\phi^a+\theta^{\alpha}\psi^a_\alpha-\theta^2 F^a$ are three real superfelds. The supersymmetric action (\ref{so3}) can be expanded is components as follows
\be
\mathcal{S}=\frac{1}{2}\int d^2x \left( -\pa^\mu\phi^a\pa_\mu\phi^a+i \psi^{a\alpha}\pa_\alpha^{\,\,\,\beta}\psi^a_\beta+F^a F^a \right).
\ee 
The supersymmetric invariant constraint in (\ref{so3}) yields to three constraints in the component fields (sum is understood in the repeated indices),
\bea
\phi^a\phi^a&=&1,\label{c1}\\
\phi^a\psi^a_\alpha&=&0,\label{c2}\\
F^a\phi^a&=&\frac{1}{2}\psi^{a\alpha}\psi^a_\alpha.\label{c3}
\eea
Taking into account (\ref{c1})-(\ref{c3}) we can eliminate the auxiliary fields from the action ($F^k=\frac{1}{2}\phi^k\psi^{a\alpha}\psi^a_\alpha $). The resulting on-shell action can be written as follows
\be
\mathcal{S}=\frac{1}{2}\int d^2x \left( -\pa^\mu\phi^a\pa_\mu\phi^a+i \psi^{a\alpha}\pa_\alpha^{\,\,\,\beta}\psi^a_\beta+\frac{1}{8}\left(\psi^{a\alpha}\psi^a_\alpha\right)^2 \right).
\ee 
This expression is reasonably simple, but we have to take into account the constraints (\ref{c1}) and (\ref{c2}). We can instead, solve explicitly the constraints (\ref{c1})-(\ref{c3}) and rewrite the action in terms of a complex superfield. To proceed, we can use a SUSY analogous of the stereographic projection
 \be
 \Phi^a=\frac{1}{(1+\Phi^\dagger \Phi)}\left(\Phi+\Phi^\dagger,-i(\Phi-\Phi^\dagger),1-\Phi^\dagger\Phi\right)\label{stereo}
 \ee 
 where $\Phi$ is a complex scalar superfield. We get in components
 \bea
\phi^a&=&\frac{1}{(1+\bar{\phi} \phi)} \left(\phi+\bar{\phi},-i (\phi-\bar{\phi}),1-\bar{\phi}\phi\right)\label{phia}\\
\psi^1_\alpha&=&\frac{1}{(1+\bar{\phi}\phi)}\left(\psi_\alpha+\bar{\psi}_\alpha-\frac{\phi+\bar{\phi}}{(1+\bar{\phi}\phi)}\chi_\alpha\right)	\label{psi1}\\
\psi^2_\alpha&=&\frac{1}{(1+\bar{\phi}\phi)}\left(-i(\psi_\alpha-\bar{\psi}_\alpha)+i\frac{\phi-\bar{\phi}}{(1+\bar{\phi}\phi)^2}\chi_\alpha\right)\label{psi2}\\
\psi^3_\alpha&=&\frac{1}{(1+\bar{\phi}\phi)}\left(-\chi_\alpha-\frac{1-\phi\bar{\phi}}{(1+\bar{\phi}\phi)}\chi_\alpha\right)\label{psi3}\\
F^1&=&\frac{F+\bar{F}}{(1+\bar{\phi}\phi)}-(\phi+\bar{\phi})H-\frac{1}{(1+\bar{\phi}\phi)^2}\chi^\alpha(\psi_\alpha+\bar{\psi}_\alpha) \label{F1}\\
F^2&=&-i\frac{F-\bar{F}}{(1+\bar{\phi}\phi)}+i(\phi-\bar{\phi})H+\frac{i}{(1+\bar{\phi}\phi)^2}\chi^\alpha(\psi_\alpha-\bar{\psi}_\alpha) \label{F2} \\
F^3&=&-\frac{F\bar{\phi}+\bar{F}\phi+\bar{\psi}^\alpha\psi_\alpha}{(1+\bar{\phi}\phi)}-(1-\phi\bar{\phi})H+\frac{1}{(1+\bar{\phi}\phi)^2}\chi^\alpha\chi_\alpha  \label{F3}
 \eea
where
\bea
\chi_\alpha&=&\bar{\psi}_\alpha \phi+\psi_\alpha \phi\\
H&=& \frac{1}{(1+\bar{\phi}\phi)^2}\left(F\bar{\phi}+\bar{F}\phi+\bar{\psi}^\alpha \psi_\alpha\right)-\frac{1}{(1+\bar{\phi}\phi)^3}\chi^\alpha\chi_\alpha.
\eea
It is easy to verify that, in terms of the new complex fields the constraints (\ref{c1})-(\ref{c3}) are automatically satisfied. If we substitute (\ref{stereo}) in (\ref{so3}), we get
\be
\mathcal{S}=-\frac{1}{2}\int d^2xd^2\theta \,g(\Phi,\Phi^\dagger)D^\alpha \Phi^\dagger D_\alpha\Phi\label{sso3}
\ee
where $g(\Phi,\Phi^\dagger)=1/(1+\Phi^\dagger\Phi)^2$ is the $\mathbb{C}P^1$ metric. The action (\ref{sso3}) constitutes the $N=1$ $\mathbb{C}P^1$ formulation of the model. Standard calculations lead to the following expression
\bea
\mathcal{S}&=&-\frac{1}{2}\int d^2 x\left( g(\phi,\bar{\phi})\left(i\pa^\alpha_{\,\,\, \beta}\bar{\psi}^\beta\psi_\alpha+i \bar{\psi}^\alpha \pa_{\alpha\beta}\psi^\beta+\pa^{\alpha\beta}\bar{\phi}\pa_{\alpha\beta}\phi-2 F\bar{F}   \right)\right.\nonumber \\
&&\left.+\pa^2_{\phi\bar{\phi}} g(\phi,\bar{\phi}) \bar{\psi}^\alpha\psi_\alpha \bar{\psi}^\beta\psi_\beta-i \pa_{\bar{\phi}} g(\phi,\bar{\phi})\pa_{\gamma\alpha}\bar{\phi}\bar{\psi}^\gamma\psi^\alpha-i \pa_\phi g(\phi,\bar{\phi})\pa_{\gamma\alpha}\phi\psi^\gamma \bar{\psi}^\alpha \right.\nonumber\\
&&\left.-  \pa_{\phi} g(\phi,\bar{\phi})\bar{F}\psi^\alpha\psi_\alpha-\pa_{\bar{\phi}} g(\phi,\bar{\phi})F\bar{\psi}^\alpha \bar{\psi}_\alpha\right).\label{compc}
\eea
We can eliminate the auxiliary field from its equation of motion 
\be
F=-\frac{ \pa_{\phi} g(\phi,\bar{\phi})}{2 g(\phi,\bar{\phi})}\psi^\alpha\psi_\alpha,\quad \bar{F}=-\frac{ \pa_{\bar{\phi}} g(\phi,\bar{\phi})}{2 g(\phi,\bar{\phi})}\bar{\psi}^\alpha\bar{\psi}_\alpha\label{aux}.
\ee
After substituting (\ref{aux}) in (\ref{compc}) we can write the action in a geometrical way
\be
\mathcal{S}_{O(3)}=-\frac{1}{2}\int d^2x g(\phi,\bar{\phi})\left(\pa^{\alpha\beta}\bar{\phi}\pa_{\alpha\beta}\phi+i\psi^\alpha D_{\alpha\beta}\bar{\psi}^\beta+i \bar{\psi}^\alpha \bar{D}_{\alpha\beta}\psi^\beta\right)+ \mathcal{R}\psi^\alpha\psi_\alpha \bar{\psi}^\beta\bar{\psi}_\beta\label{aco3}
\ee
where $D_{\alpha\beta}=\pa_{\alpha\beta}-\Gamma^{\phi}_{\phi\phi}\pa_{\alpha\beta}\phi$ and $\mathcal{R}$ is the Riemann tensor for $\mathbb{C}P^1$.









\section{A bosonic twin for the $O(3)$ $\sigma$-model}

Obviously, actions (\ref{so3}) and (\ref{sso3}) are equivalent. They represent the same classical field theories but expressed by means of different target-space variables. Hence, their bosonic and fermionic sectors coincide and are related by the transformations (\ref{phia})-(\ref{F3}). 

Another question is whether it is possible to construct different, physically {\it not equivalent} super symmetric extensions of a given bosonic model. In other words, we want to construct actions which are: 1) invariant under the sypersymmetric transformations; 2) share the same bosonic sector; 3) but differ as far as the fermionic part is considered. 

This question has been answered affirmatively in the literature. For example in \cite{Queiruga1,Queiruga2,Bolognesi} different supersymmetric extensions of the baby Skyrme model were proposed. However, the fermionic part of the supersymmetric models contains potentially dangerous higher-derivative terms.  A different proposal was made in \cite{Nitta1,Nitta2} (in four dimensions) and in \cite{Queiruga3,Queiruga4,Queiruga5} (in three dimensions) with $N=1$ and $N=2$ SUSY. In these cases, the fermionic part of the non-equivalent supersymmetric extensions suffers from the appearance of derivative terms involving the auxiliary field. This may promote the auxiliary field to a dynamical one. Furthermore, all these SUSY extensions reproduce the bosonic model only on-shell, i.e., once we eliminate the auxiliary degrees of freedom from the action.

Here the aim is to analyse the possibility of the existence of non-equivalent SUSY extensions of a given (bosonic) model with an additional condition, that 4) no higher-derivative terms in the fermionic sector are allowed. 

It turns out that in 2 and 3 dimensions and with $N=1$ SUSY there is not too much freedom. Let us look for terms with trivial (empty) bosonic sector. Then, such terms can be added to a SUSY action (for example (\ref{sso3})) without any deformation of the original bosonic sector. A necessary condition for such a term is that it requires at least four odd operators (in the number of superderivatives). Moreover, the degree of each operator (the number of superderivatives) cannot be greater than one, otherwise we will generate higher-time derivatives - a possibility which we excluded from the very beginning. At the end only one combination remains
\be
\mathcal{L}^d=\int d^2\theta H(\Phi,\Phi^\dagger)D^\alpha \Phi^\dagger D_\alpha\Phi D^\beta \Phi^\dagger D_\beta \Phi\label{def1}
\ee
where the function $H(\Phi,\Phi^\dagger)$ is arbitrary and depends only on the superfields but not on derivatives. Here the target-space manifold $\mathcal{M}_T$ verifies $\dim_{\mathbb{C}}=1$. If $\dim_{\mathbb{C}}>1$ more combinations are allowed. Note that the combination $D^\alpha \Phi D_\alpha\Phi D^\beta \Phi^\dagger D_\beta \Phi^\dagger$ is proportional to (\ref{def1}) since
\be
D_\alpha\Phi D_\beta\Phi=\frac{1}{2}C_{\beta\alpha}D^\gamma\Phi D_\gamma\Phi
\ee
As we will see, the addition of these terms to a given SUSY model does not change the bosonic sector. In this sense we say that they generate ``bosonic twins''. The expansion in components of (\ref{def1}) leads to 
\bea
\mathcal{L}^d&=&\left(-H_{\bar{\phi}}\bar{F}-H_\phi F\right)\bar{\psi}\psi \bar{\psi}\psi\nonumber\\
&&+2 H  \left\{\left(i\pa^{\alpha}_{\,\,\,\beta}\bar{\psi}^\beta \psi_\alpha+i\bar{\psi}^\alpha\pa_{\alpha\beta}\psi^\beta+\pa^{\beta\alpha}\bar{\phi}\pa_{\beta\alpha}\phi-2 F\bar{F}  \right)\bar{\psi}\psi\right.\nonumber \\
&&\left. +\left(\left(-\frac{1}{4}\pa^{\beta\alpha}\bar{\phi}\pa_{\beta\alpha}\bar{\phi}+\frac{1}{2}\bar{F}^2\right)\psi\psi+\text{h.c.} \right)\right.\nonumber \\
&&\left. +F\bar{F} \bar{\psi}\psi+\left(-i\pa^{\beta\alpha}\bar{\phi}F\psi_\alpha\bar{\psi}_\beta+\text{h.c.}\right)+\pa^{\gamma\alpha}\phi\pa_{\gamma\beta}\bar{\phi}\bar{\psi}_\alpha\psi^\beta\right\}\label{defcomp}
\eea
and, as expected, no higher-derivatives appear in the action and $\mathcal{L}^d\vert_{\psi=0}=0$. It is important to note that the fact that the superfield $\Phi^\dagger$ is complex is crucial - otherwise these new terms vanish. This result in the following observation: this construction of the twins trivialize if we consider a single real scalar bosonic model. 

Specifically, we take the $O(3)$ model in the $\mathbb{C}P^1$ formulation (\ref{sso3}) and add (\ref{defcomp}). Then we obtain a new model sharing the off-shell bosonic sector with the original one. Let us analyze the on-shell action in detail. We first need to eliminate the auxiliary field $F$ from (\ref{sso3})+(\ref{defcomp}). We obtain
\be
F= -\frac{g_\phi}{2g}\psi\psi+\frac{H}{2g}i\pa^{\alpha\beta}\phi\bar{\psi}_\alpha\psi_\beta-\frac{H_{\bar{\phi}}}{2g}(\bar{\psi}\psi)^2-\frac{H g_{\bar{\phi}}}{2g^2}(\bar{\psi}\psi)^2\label{auxdef}.
\ee
The first term in (\ref{auxdef}) corresponds to the original $O(3)$ model while the others are originated by (\ref{defcomp}). After substituting (\ref{auxdef}) in (\ref{sso3})+(\ref{defcomp}) we get
\bea
\mathcal{L}_{O(3)}^d&=& \frac{1}{2}g(\phi,\bar{\phi})\left(-2\pa^{\mu}\bar{\phi}\pa_{\mu}\phi+i\psi^\alpha D_{\alpha\beta}\bar{\psi}^\beta+i \bar{\psi}^\alpha \bar{D}_{\alpha\beta}\psi^\beta\right)\nonumber\\
&-&H(\phi,\bar{\phi})\left(\pa_\mu \bar{\phi}\pa^\mu\bar{\phi}\psi\psi+\pa_\mu \phi\pa^\mu\phi\bar{\psi }\bar{\psi}-2\pa^{\gamma\alpha}\phi \pa_{\gamma\beta}\bar{\phi}\psi_\alpha\bar{\psi}^\beta\right)+... \label{aco3}
\eea
where the dots stand for quartic fermionic terms. The first line in (\ref{aco3}) corresponds to the original SUSY $\mathbb{C}P^1$ model, while the second one is originated from the deformation (\ref{defcomp}). 

\section{Adding a potential}

It is a well-known fact that in two dimensions one can add a prepotential term to the action without spoiling the $N=1$ SUSY. For future purposes we will restrict the potential to be holomorphic and antiholomorphic functions of the superfields. Then, a general SUSY non-linear $\sigma$-model on complex one-dimensional manifolds (fixed by a particular choice of the metric function $g$) reads
\bea
S_\sigma=-\frac{1}{2}\int d^2x d^2\theta g (\Phi,\Phi^\dagger)D^\alpha \Phi^\dagger D_\alpha \Phi\label{sigma}, \\
S_P=\int d^2x d^2\theta\left(W(\Phi)+W(\Phi^\dagger)\right)\label{pot}
\eea
where $W(\Phi)$ defines the prepotential part. Expanding the potential term in components we find
 \be
 S_P=\int d^2x\left(W'(\phi)F+\frac{1}{2}W''(\phi)\psi^\alpha\psi_\alpha +\text{h.c.} \right).
\ee

Let us start with the purely bosonic sector. The static energy functional  can be written as 
\be
E=\int dx \left(g \phi' \bar{\phi}'+\frac{W'\bar{W}'}{g}\right)\label{ene1}
\ee
where $\phi'\equiv\pa_x \phi$ and $W'\equiv\pa_\phi W$. We can use the Bogomolny trick and rewrite the energy integral by completing the square
\be
E=\int dx \left(\sqrt{g}\phi'-e^{i\alpha}\frac{\bar{W}'}{\sqrt{g}}\right)\left(\sqrt{g}\bar{\phi}'-e^{-i\alpha}\frac{W'}{\sqrt{g}}\right)+2\text{Re}\lbrack e^{-i\alpha}\phi'W'\rbrack.
\ee
Note that $\alpha$ is an arbitrary quantity. We obtain the strongest lower bound for the energy for $\alpha=\text{arg}(\phi'W')$, therefore
\be
E\geq 2\int d x\vert \phi' \bar{W}'\vert.
\ee
Obviously, the bound in saturated when the field obeys the Bogomolny equation (where we assume that the prepotential allows for static solitonic solutions)
\be
\phi'=e^{i\alpha}\frac{\bar{W}'}{g}\label{bog}.
\ee
It is easy to verify that the solutions of (\ref{bog}) obey the second order equation from (\ref{sigma}). 

Now we will analyze the fermionic sector. After eliminating the auxiliary field from the action (\ref{sigma})+(\ref{pot}) we get
\bea
\mathcal{L}_{\sigma P}&=&g\left(-\pa_\mu \phi\pa^\mu\bar{\phi}-\frac{i}{2}\pa^{\alpha}_{\,\,\,\beta}\bar{\psi}^\beta\psi_\alpha-\frac{i}{2}\bar{\psi}^\alpha \pa_{\alpha\beta}\psi^\beta\right)-\frac{W'\bar{W}'}{g}+\nonumber\\
&+&\frac{1}{2}W''\psi^\alpha \psi_\alpha+\frac{1}{2}\bar{W}''\bar{\psi}^\alpha\bar{\psi}_\alpha-W'\frac{g_\phi}{2g}\psi^\alpha\psi_\alpha-\bar{W}'\frac{g_{\bar{\phi}}}{2g}\bar{\psi}^\alpha\bar{\psi}_\alpha+\nonumber\\
&+&\frac{i}{2}g_\phi \psi^{\alpha}\bar{\psi}^\beta\pa_{\alpha\beta}\phi-\frac{i}{2}g_{\bar{\phi}}\bar{\psi}^\alpha\psi_\beta\pa_{\alpha}^{\,\,\,\beta}\bar{\phi}\label{full}+\mathcal{O}(\psi^2\bar{\psi}^2).
\eea
In the next step we explicitly express the spinors in chiral components $(\psi_+,\psi_-)$. The static fermionic part of the Lagrangian in these new variables takes the following form
\bea
L_{\sigma P}^f&=&-i\frac{g}{2}\bar{\psi_+}\psi_+'+i\frac{g}{2}\bar{\psi_+}'\psi_++i\frac{g}{2}\bar{\psi_-}\psi_-'-i\frac{g}{2}\bar{\psi_-}'\psi_-\nonumber\\
&+&i W'\frac{g_\phi}{g}\psi_+\psi_-+i\bar{W}'\frac{g_{\bar{\phi}}}{g}\bar{\psi}_+\bar{\psi}_--i W'' \psi_+\psi_--i \bar{W}''\bar{\psi}_+\bar{\psi}_-\nonumber\\
&-&\frac{i}{2}g_\phi \psi_-\bar{\psi}_-\phi'+\frac{i}{2}g_\phi \psi_+\bar{\psi}_+\phi'-\frac{i}{2}g_{\bar{\phi}}\bar{\psi}_-\psi_-\bar{\phi}'+\frac{i}{2}g_{\bar{\phi}}\bar{\psi}_+\psi_+\bar{\phi}'+\mathcal{O}(\psi^2\bar{\psi}^2)\label{ferm}
\eea
Here, two observations are in order. First, the Lagrangian is invariant under the $N=1$ supersymmetry transformations
\bea
\delta\phi&=&-\epsilon^\alpha \psi_\alpha\label{t1} \\
\delta \psi_\alpha&=&-\epsilon^\beta\left(C_{\alpha\beta}F-i\pa_{\alpha\beta}\phi\right)\label{t2}\\
\delta F&=&-\epsilon^\alpha i\pa_{\alpha}^{\,\,\,\beta}\psi_\beta\label{t3}
\eea
where $\epsilon$ is a real spinor. Second, the requirement that the theory is invariant under the $N=2$ supersymmetry can be achieved by promoting $\epsilon$ to a complex object. This is equivalent to say that we have the transformations (\ref{t1})-(\ref{t3}) with real parameter followed by a phase rotation for the fermions. In terms of the chiral components in the Lagrangian (\ref{ferm}) it means 
\be
\psi_{\pm}\rightarrow e^{\pm i\alpha}\psi_{\pm},\quad \bar{\psi}_{\pm}\rightarrow e^{\mp i\alpha}\psi_{\pm}\label{ext}.
\ee
 The substitution of (\ref{ext}) in (\ref{ferm}) leaves the Lagrangian invariant implying that the model has an extra supersymmetry. This can be confirmed directly by rewriting (\ref{full}) in the $N=2$ SUSY language. Namely, 
\be
\mathcal{L}_{N=2}=\int d^2 \bar{\theta} d^2\theta K(\Phi^\dagger,\Phi)+\left(\int d^2\theta W(\Phi)+\text{h.c.}\right)
\ee
where $K$ is the K\"{a}hler potential. Now, we will consider the fermionic (zero-mode) equation obtained from (\ref{ferm})
\be
 \left(\begin{matrix}
    g\pa_x+g_\phi \phi'  \quad     & \bar{W}''-\frac{g_{\bar{\phi}}}{g}\bar{W}'\\
    W''-\frac{g_\phi}{g} W' \quad        & g\pa_x+g_{\bar{\phi }}\bar{\phi}' 
\end{matrix}\right)\left(\begin{matrix}\psi_+\\ \bar{\psi}_- \end{matrix} \right)=0\label{fereq}
\ee
On the other hand from (\ref{t2}) we find
\be
\left(\begin{matrix}\delta\psi_+ \\ \delta \bar{\psi}_- \end{matrix}\right)=-\frac{i}{2}\left( \begin{matrix}F+e^{-i\alpha}\phi' & F-e^{-i\alpha}\phi'\\  
-\bar{\phi}'-e^{-i\alpha}\bar{F} & e^{-i\alpha} \bar{F}-\bar{\phi}' \end{matrix}\right)\left(\begin{matrix}\eta \\ \xi\end{matrix}\right).\label{zero}
\ee
where 
\be
\eta=\frac{1}{2}\left(\epsilon^2+e^{i\alpha}\epsilon^1\right),\quad \xi=\frac{1}{2}\left(\epsilon^2-e^{i\alpha}\epsilon^1\right)
\ee
It is straightforward to see that  solutions of the Bogomolny equation (\ref{bog}) are preserved by supersymmetry transformations with $\eta=0, \xi\neq 0$. This has a consequence that the zero-mode equation (\ref{fereq}) is automatically satisfied for
 \be
 \psi_+=-i e^{-i\alpha}\phi'\eta,\quad \bar{\psi}_-=i\bar{\phi}'\eta
 \ee
 Therefore, the fermions are parametrized by only one real constant $\eta$ (see for example \cite{Townsend}). As a consequences only $1/2$ of supersymmetry is preserved (in $N=1$). If the theory has a hidden extended supersymmetry only $1/4$ of the ($N=2$) supersymmetry is preserved. The connection between solitons ad fermionic zero-modes in SUSY theories has been extensively discussed in the literature \cite{diVecchia,Rossi}.





\section{The bosonic twin and fermion zero-modes}

As we have seen before, the introduced deformation term does not modify the bosonic sector of the original action while it nontrivially contributes to the fermionic sector even at quadratic order. Furthermore, the auxiliary field also gets contributions at second order in the spinors 
\be
F=-\frac{W'}{g}-2i \frac{H}{g}\pa^{\alpha\beta}\bar{\phi}\psi_\alpha\bar{\psi}_\beta-2H\frac{\bar{W}'}{g^2}\bar{\psi}^\alpha\psi_\alpha+2 H \frac{W'}{g}\psi^\alpha\psi_\alpha-\frac{g_\phi}{2g}\psi^\alpha\psi_\alpha+\mathcal{O}(\psi^2\bar{\psi}^2)
\ee
 After eliminating the auxiliary field, the full $O(3)$ sigma model action with the potential and the deformation term included is
\bea
\mathcal{L}_{\sigma P d}&=&g\left(-\pa_\mu \phi\pa^\mu\bar{\phi}-\frac{i}{2}\pa^{\alpha}_{\,\,\,\beta}\bar{\psi}^\beta\psi_\alpha-\frac{i}{2}\bar{\psi}^\alpha \pa_{\alpha\beta}\psi^\beta\right)-\frac{W'\bar{W}'}{g}+\nonumber\\
&&+\frac{1}{2}W''\psi^\alpha \psi_\alpha+\frac{1}{2}\bar{W}''\bar{\psi}^\alpha\bar{\psi}_\alpha-W'\frac{g_\phi}{g}\psi^\alpha\psi_\alpha-\bar{W}'\frac{g_{\bar{\phi}}}{g}\bar{\psi}^\alpha\bar{\psi}_\alpha+\nonumber\\
&&+\frac{i}{2}g_\phi \psi^{\alpha}\bar{\psi}^\beta\pa_{\alpha\beta}\phi-\frac{i}{2}g_{\bar{\phi}}\bar{\psi}^\alpha\psi_\beta\pa_{\alpha}^{\,\,\,\beta}\bar{\phi}\nonumber\\
&&+ H \left(\frac{W'^2}{g^2}\psi^\alpha \psi_\alpha+\frac{\bar{W}'^2}{g^2}\bar{\psi}^\alpha\bar{\psi}_\alpha-2\frac{W'\bar{W}'^2}{g^2}\bar{\psi}^\alpha\psi_\alpha+2i\frac{\bar{W}'}{g}\pa^{\alpha\beta}\psi_\alpha\bar{\psi}_\beta\nonumber \right.\\
&&\left. -2i\frac{W'}{g}\pa^{\alpha\beta}\psi_\alpha\bar{\psi}_\beta+4\pa^\mu\bar{\phi}\pa_\mu\phi \bar{\psi}^\alpha\psi_\alpha-\pa^\mu\phi\pa_\mu\phi\bar{\psi}^\alpha\bar{\psi}_\alpha\right.\nonumber\\
&&\left.-\pa^\mu\bar{\phi}\pa_\mu\bar{\phi}\psi^\alpha\psi_\alpha+2\pa^{\gamma\alpha}\phi\pa_{\gamma\beta}\bar{\phi}\psi_\alpha\bar{\psi}^\beta  \right)+\mathcal{O}(\psi^2\bar{\psi}^2)
\label{fulld}
\eea
We write the Lagrangian (\ref{fulld}) in terms of the chiral spinors $(\psi_+,\psi_-)$ using the following replacements
\bea
\psi^\alpha\psi_\alpha &\rightarrow & 2i\psi_+ \psi_- \\
\bar{\psi}^\alpha\bar{\psi}_\alpha &\rightarrow & 2i\bar{\psi}_+ \bar{\psi}_- \\
\bar{\psi}^\alpha\psi_\alpha  &\rightarrow & i\left(\bar{\psi}_-\psi_+-\bar{\psi}_+\psi_-\right)\label{vioferm}
\eea
 
The invariance under $N=2$ SUSY requires that fermion-number terms are absent from the action. However, the deformation introduces two such terms: one proportional to $\bar{\psi}^\alpha\psi_\alpha$ and the last term in (\ref{fulld}) proportional to $\pa^{\gamma\alpha}\phi\pa_{\gamma\beta}\bar{\phi}\psi_\alpha\bar{\psi}^\beta$. This means that they do not respect the symmetry (\ref{ext}). As a consequence, the deformation term breaks the original $N=2$ of the $\sigma$-model to $N=1$ SUSY. 

Although by construction our deformation prescription does not modify the bosonic sector, which in particular means that the BPS sector remains unchanged, the linearized equation for the fermions contains a nontrivial contribution proportional to $H$ (the arbitrary function in the deformation part). As we explained in the previous section the fermionic zero-modes are connected with the BPS sector via the supersymmetric transformations. Therefore, one expects a relation between the fermionic zero-mode equations in deformed models (different $H$). Indeed, we find that in the background of a solution satisfying (\ref{bog}) the zero-mode equation including the deformation term reduces to (\ref{fereq}). Moreover, the fermion-number violating terms are effectively absent from the action, recovering in some sense the original $N=2$ supersymmetry. To prove this we consider the zero mode equation in the deformed model
 \bea
 &&\left(g\pa_x+g_\phi\phi'-2H \left(\frac{\bar{W}'}{g}\bar{\phi}'-\frac{W'}{g}\phi'\right)\right)\psi_+\nonumber\\
 &&+\left(\bar{W}''-\bar{W}'\frac{g_\phi}{g}+2H \left(\phi'^2-\frac{\bar{W}'^2}{g^2}\right)\right)\bar{\psi}_-\nonumber\\
 &&-2H\left(\phi'\bar{\phi}'-\frac{W'\bar{W}'}{g^2}\right)\psi_-=0\label{deffer1}.
 \eea
Obviously, all terms proportional to $H$ vanish in the Bogomolny sector, that is, when $\phi'=\frac{\bar{W}'}{g}$ leading to the equation (\ref{fereq}). Note also that, eq. (\ref{deffer1}) is equivalent to (\ref{fereq}) only for $\alpha=0$, i.e. not for the continuous family of BPS equations. This is because the deformation term breaks the symmetry $\phi\rightarrow e^{i\alpha}\phi$ which is present in the original model.
 

 \section{The $O(3)$ $\sigma$-model in $2+1$ dimensions}
 
 In this section we consider the $O(3)$ $\sigma$-model in $2+1$ dimensions. This obviously has a nontrivial impact on the solitonic structure of the model. First of all, static solutions can be treated as maps $\vec{\phi} : \mathbb{R}^2 \cup \{ \infty \} \cong \mathbb{S}^2 \rightarrow \mathbb{S}^2$, where a single asymptotic value of the bosonic field is assumed. This allows for a one point compactification of the base space. Now, maps between two-spheres are classified by a pertinent topological index 
\be
\pi_2(\mathbb{S}^2)=\mathbb{Z}.
\ee 
Secondly, due to the Derrick's theorem there is no stable, finite energy static solutions if a potential term is present. 
Note, that one can keep the potential part if another (higher-derivative term) is simultaneously introduced. This will be analyzed in the next sections. Once we neglect the potential, the resulting pure $O(3)$ model is a BPS theory.  This means that the static energy functional is bounded from below by the topological charge
 \be
 E\geq  4\int d^2x\frac{\vert \vert \pa_{\bar{z}}\phi\vert^2-\vert\pa_z\phi \vert^2 \vert}{(1+\phi\bar\phi)^2} 
 \ee 
 while the bound is saturated for solutions of the Bogomolny equations, which in this case are just  the Cauchy-Riemann equations
 \be
 \pa_x \phi=\pm i\pa_y \phi,\quad  \pa_x \bar{\phi}=\mp i\pa_y \bar{\phi}\label{cauchy}.
 \ee
Hence, the BPS sector (solutions of the Bogomolny equations) is constituted by holomorphic/antiholomorphic functions 
 \be
 \phi=\phi(z),\,\, z=x+iy \;\;\;\; \text{or} \;\;\;\; \phi=\phi(\bar{z}),\,\, \bar{z}=x-iy.
 \ee
 
The fermionic zero-modes in the BPS sector can be obtained using the supersymmetric transformation for the fermionic degrees of freedom. From (\ref{t2}) we have
\be
\left(\begin{matrix}\delta\psi_+ \\ \delta\psi_- \end{matrix}\right)=\left( \begin{matrix}-i\pa_{\bar{z}}\phi & -i \pa_z\phi\\  
-\pa_{\bar{z}}\phi &  \pa_z\phi \end{matrix}\right)\left(\begin{matrix}\eta \\ \xi\end{matrix}\right)\label{zero2}
\ee
where 
\be
\eta=\frac{1}{2}\left(\epsilon^1-i\epsilon^2\right),\quad \xi=\frac{1}{2}\left(\epsilon^1+i\epsilon^2\right).
\ee 
The Dirac equation from (\ref{aco3}) can be written as follows
\be
g\left(\pa_z \chi+\pa_{\bar{z}}\chi^c\right)+\pa_\phi g\left(\pa_z\phi\chi+\pa_{\bar{z}}\phi\chi^c\right)=0\label{fer2}
\ee
where $\chi=\psi_++i\psi_-$ and $\chi^c=\psi_+-i\psi_-$. Like in the one-dimensional case, in the background of a BPS solution the fermions depends only on one real parameter. Let us take $\pa_z\phi=0$, we have from (\ref{zero2})
\be
\chi=-2i\pa_{\bar{z}}\phi\eta,\quad \chi^c=0.
\ee
These solutions automatically satisfy (\ref{fer2}) since $\pa_{\bar{z}}\pa_z\phi=0$. Due to the supersymmetry of the model we can express the fermionic zero-modes through holomorphic/antiholomorphic derivatives of the BPS solutions. 

Now, let us see what happens if we add the deformation terms. Of course, as in the one-dimensional case the BPS sector does not change. On the other hand, the linearized fermionic equation gets extra terms
\bea
&&g\left(\pa_z \chi+\pa_{\bar{z}}\chi^c\right)+\pa_\phi g\left(\pa_z\phi\chi+\pa_{\bar{z}}\phi\chi^c\right)-\frac{H}{2}(\vert\pa_{\bar{z}}\phi\vert^2+\vert\pa_z\phi\vert^2)(\chi-\chi^c)\nonumber\\
&&-H\pa_z\phi \pa_{\bar{z}}\phi(\bar{\chi}-\bar{\chi}^c)+\frac{H}{2}(\vert\pa_{\bar{z}}\phi\vert^2-\vert\pa_z\phi\vert^2)(\chi+\chi^c)=0\label{fer3}.
\eea

However, in the background of BPS solutions, for example $\pa_z\phi=0\Rightarrow \chi^c=0$,  solutions of (\ref{fer3}) are of the form
\be
\chi=-2i\pa_{\bar{z}}\phi\eta.
\ee

Therefore we observe the same effect. In the background of the BPS solutions the fermionic zero-modes are independent of the deformation terms. Besides, all fermion-number violating terms in the action are absent once we substitute the BPS equation (\ref{cauchy}), implying an ``on-shell" restoration of the $N=2$ SUSY.


\subsection{The quartic fermionic terms}

The analysis of the previous section was based on the linearized fermionic sector, where we did not take into account the higher fermionic contributions. In the original $\sigma$-model only a quartic term appears accompanying the Riemann tensor of the target space manifold. Once we add the deformation term the situation is more complicated. The quartic action in fermions for the pure deformed $\sigma$-model can be written as 
\be
\mathcal{L}_4=H\left(i\pa_{\alpha\beta}\bar{\psi}^\beta\bar{\psi}^\alpha \psi^2+i\pa_{\alpha\beta}\psi^\beta\psi^\alpha\bar{\psi}^2\right)+\left(\mathcal{R}-2H^2\frac{\pa_\mu\bar{\phi}\pa^\mu\phi}{g}\right)\bar{\psi}^2\psi^2\label{quar}.
\ee

In the pure undeformed $\sigma$-model the term proportional to $\mathcal{R}$ cannot be eliminated unless the target manifold is trivial. In our case, extra fermionic terms involving derivatives appear and cannot be eliminated unless $H=0$ (the first two terms in (\ref{quar})). On the contrary,  for a special choice of the function $H$, the term proportional to $\bar{\psi}^2\psi^2$ can be eliminated in the background of the BPS solutions. Let us take, for example the holomorphic solution $\pa_{\bar{z}}\phi=0$. The last term in (\ref{quar}) ca be rewritten as
\be
\left(\mathcal{R}-H^2\frac{\vert\pa_z\phi\vert^2}{g}\right)\bar{\psi}^2\psi^2\label{Riem}.
\ee
Let F(z) be a holomorphic function such that $ F(z)=\pa_z\phi(z)$, where $\phi(z)$ is a particular BPS solution  $\phi=f(z)$. Now $F(z)$ defined as the holomorphic derivative of a particular solution $\phi(z)$ can be written as a function of $\phi$ itself for \textit{this particular solution}
\be
F(z)=\pa_z \phi(z)\Rightarrow F(\phi)=f'(f^{-1}(\phi))
\ee
where the prime indicates differentiation with respect to its argument. The choice
\be
H^2=\frac{\mathcal{R}g}{F(\phi)F(\bar{\phi})}
\ee 
eliminates (\ref{Riem}) and leads to the following quartic contribution
\be
\mathcal{L}_4\vert_{\text{BPS}}=\frac{\mathcal{R}g}{F(\phi)F(\bar{\phi})}\left(i\pa_{\alpha\beta}\bar{\psi}^\beta\bar{\psi}^\alpha \psi^2+i\pa_{\alpha\beta}\psi^\beta\psi^\alpha\bar{\psi}^2\right).
\ee

We want to emphasize the fact that the quartic term (\ref{Riem}) is only eliminated in the background of the BPS solution for which H was constructed.









\section{higher-derivative terms in the $N=2$ bosonic twins}

We can work directly in the $N=2$ language to suggest why $N=2$ bosonic twins cannot exist. Let us start with the $N=2$ version of the $O(3)$ $\sigma$-model (or more precisely the $\mathbb{C}P^1\, \sigma$-model). We have
\be
\mathcal{L}_{\mathbb{C}P^1}^{N=2}=\int d^2\theta d^2\bar{\theta}\log \left(1+\Phi^\dagger \Phi\right)
\ee

We need first to saturated the Grassmann integration to generate a term without bosonic sector. The key term is given now by the fourth derivative term
\be
\mathcal{L}_4=\int d^2\theta d^2\bar{\theta} D^\alpha \Phi \bar{D}^{\dot{\beta}}\Phi^\dagger D_\alpha \Phi \bar{D}_{\dot{\beta}}\Phi^\dagger.\label{quart}
\ee

By multiplying this term with fermionic objects we could in principle construct pure fermionic actions, and as we did before, we could construct inequivalent SUSY extensions of a given bosonic model. In this case there are two terms verifying the properties described above
\bea
T_6&=&\Re e\,D^\alpha \Phi \bar{D}^{\dot{\beta}}\Phi^\dagger D_\alpha \Phi \bar{D}_{\dot{\beta}}\Phi^\dagger  D^\beta \Phi D_\beta \Phi\label{t6}\\
T_8&=&\left(D^\alpha \Phi \bar{D}^{\dot{\beta}}\Phi^\dagger D_\alpha \Phi \bar{D}_{\dot{\beta}}\Phi^\dagger\right)^2\label{t8}.
\eea

But now the Lagrangian (\ref{quart}) possesses nontrivial bosonic sector, namely
\be
\mathcal{L}_4\propto\left((\pa_\mu \phi)^2(\pa_\nu\bar{\phi})^2-2 \bar{F}F\pa_\mu\phi \pa^\mu\bar{\phi}+(F\bar{F})^2  \right)\label{T8}
\ee
and  $T_6$ and $T_8$ verify 
\be
T_6\vert_{\theta,\bar{\theta}=0}=T_8\vert_{\theta,\bar{\theta}=0}=0.
\ee

 This leads to $T_6=T_8=0$ and therefore, $N=2$ SUSY does not allow for the extra extensions if the dimension of the target space manifold ($\mathcal{M}_T$)  is two (or in other words, in the target manifold can be constructed in terms of only one chiral superfield). Since, in order to have $N=2$ SUSY, $ \mathcal{M}_T$ must be K{\"a}hler and therefore $\dim \mathcal{M}_T\neq 3$. This implies that $\mathcal{M}_T$ must  be K{\"a}hler and $\dim \mathcal{M}_T\geq 4$. The first consequence is that terms of the form (\ref{def1}) cannot be extended to $N=2$ (as we verified explicitly in the previous section). Let us assume that we have a $N=1$ model built in terms of two complex superfields (i.e. $\dim\mathcal{M}_T=4$). We can construct $N=2$ pure fermionic terms is this case. It is possible to generate four different terms with six derivatives and six superfields, namely
\bea
&&\left( D\Phi^1\right)^2\left(\bar{D}\Phi^{1\dagger}\right)^2 \left(D\Phi^2\right)^2+\text{h.c}\\
&&\left( D\Phi^1\right)^2\left(\bar{D}\Phi^{2\dagger}\right)^2 \left(D\Phi^2\right)^2+\text{h.c}\\
&&\left( D\Phi^1 D\Phi^2 \right )^2\left(\bar{D}\Phi^{1\dagger}\right)^2+\text{h.c}\\
&&\left( D\Phi^1 D\Phi^2 \right )^2\left(\bar{D}\Phi^{2\dagger}\right)^2+\text{h.c}
\eea  
and three different terms involving eight derivatives and eight superfields
\bea
&&\left( D\Phi^1\right)^2\left(\bar{D}\Phi^{1\dagger}\right)^2\left( D\Phi^2\right)^2\left(\bar{D}\Phi^{2\dagger}\right)^2\\
&&\vert D\Phi^1 D\Phi^2\vert^4\\
&&\left( D\Phi^1 D\Phi^2 \right )^2\left(\bar{D}\Phi^{1\dagger}\right)^2\left(\bar{D}\Phi^{2\dagger}\right)^2+\text{h.c}
\eea
where $ \left( D\Phi\right)^2= D^\alpha \Phi D_\alpha \Phi$, etc. After expanding these terms in components we get
\bea
&&\frac{i}{2}\psi^i \sigma^\mu \bar{\psi}^j\left(\pa_\mu \bar{\phi}^i \square \phi^j-\pa_\mu \phi^i\square\bar{\phi}^j\right)\bar{\psi}^i \psi^j\subset \mathcal{L}_6\\
&&\frac{i}{2}\psi^i \sigma^\mu \bar{\psi}^j\left(\pa_\mu \bar{\phi}^i \square \phi^j-\pa_\mu \phi^i\square\bar{\phi}^j\right)\bar{\psi}^i \psi^j\bar{\psi}^i \psi^j\subset \mathcal{L}_8.
\eea 

Thus, bosonic twin cannot exist with $N=2$ SUSY unless we allow for higher-derivative terms in the fermionic sector.

\section{Summary}

In this work we have constructed new supersymmetric versions of the nonlinear $\sigma$-model with two-dimensional target-space manifold. This construction is based on the addition of a pure fermionic term (supersymmetric invariant and with vanishing bosonic sector) which is independent of the model apart from its field content. After the expansion in components we have shown that the deformation term is well-behaved in the sense that it does not contain higher-derivative terms. If the absence of higher-derivative contributions is imposed it turns out that such a term is unique (for $\dim \mathcal{M}_T=2$) up to an overall function depending on the superfields but not on derivatives. Besides, it does not contain derivatives acting on the auxiliary field (although the appearance of the derivative of the auxiliary field not always implies that $F$ becomes dynamical - it leads usually to the generation of higher-derivative terms for the physical fields \cite{Queiruga5}).

The inclusion of the deformation term to the SUSY nonlinear $\sigma$-model has the two main effects:
\begin{enumerate}
\item Since it does not modify the bosonic sector the new action constitutes a new supersymmetric extension of the original $\sigma$-model with properly deformed fermionic sector.
\item The deformation term is strictly $N=1$, i.e. there is no hidden extended supersymmetry. This implies that our supersymmetric versions of the $\sigma$-model are strictly $N=1$ (note that the usual formulations of the $\sigma$-model are implicitly or explicitly $N=2$).
\end{enumerate}

These properties lead to interesting consequences. First of all, it follows trivially from the preservation of the bosonic part that the BPS sector is also not modified. On the other hand, the fermionic sector receives nontrivial contributions even at the quadratic order. This might suggest that the relationship between BPS solutions and fermionic zero-modes is broken by the deformation term. However, as we proved in the paper, everything combines is such a way that the fermionic zero-mode equation remains unaltered (w.r.t. the original $\sigma$-model) in the background of the BPS solutions. At the same time, all fermion-number violating terms generated by the deformation disappear leading to an ``on-shell restoration" of the original $N=2$ SUSY. Needless to say that, for general bosonic solutions (non BPS) the solutions of the Dirac equation in the deformed model have no relation with the original solutions (and the fermion-number violating terms are not eliminated).

Secondly, the standard SUSY $\sigma$-model contains a quartic fermionic coupling proportional to the Riemann tensor on $\mathcal{M}_T$. This term cannot be eliminated in a supersymmetric invariant way unless $\mathcal{M}_T$ is flat. If we add the deformation part then new quartic fermionic terms show up. They contain a nontrivial contribution from the bosonic field as well. This can lead to an effective elimination of this quartic term, in the background of a BPS solution, if an appropriate choice of the arbitrary function $H$ is made. 
Nonetheless, the quartic fermionic part does not trivialize since a derivative fermionic term remains. For $\sigma$-models admitting a potential (for example in $1+1$ dimensions) the elimination of the Riemann tensor term can be achieved by a suitable choice of the prepotential.

Thirdly, the deformation term can be added to {\it any} supersymmetric version of a model based on the same bosonic d.o.f. as $O(3)$ $\sigma$-model i.e., a three component unit iso-vector $\vec{\phi}$. This concerns for example the baby Skyrme model \cite{baby} and the BPS baby Skyrme model \cite{bBPS} (a lower-dimensional counterpart of the BPS Skyrme model \cite{BPS}). Therefore our construction provides a new set of well behaved $N=1$ SUSY versions of these topologically nontrivial models \cite{Queiruga1}-\cite{Queiruga3}. It would be very desirable to analyzed the fermionic sector of these new extensions in detail with a particular focus on the fermionic zero-modes, however, this issue is beyond thee scope of the paper.

Finally, we have analyzed the bosonic twins with extended supersymmetry. It turns out that if one imposes the higher-derivative restriction such models cannot exist. Moreover, using the connection between $N=1$ SUSY in four dimensions and extended SUSY in three dimension (via dimensional reduction), one can lift this non-existence to the four-dimensional case.
 
\vspace*{0.2cm} There are several straightforward directions in which presented analysis can be further investigated. 

As we have already noticed our deformation applies for any field theory with the unit, three component iso-vector field. It would be interesting to present a complete and systematic classification of {\it all} supersymmetric extensions of the pertinent bosonic models ($O(3)$ model, the baby Skyrme and the baby BPS Skyrme models) also with the case when high-derivative terms are taken into account \cite{Bolognesi}. 

Another possibility is to repeat our construction for models with higher-dimensional target-space and (or) in higher dimensions. This would include supersymmetric $O(n)$ $\sigma$-model and especially the Skyrme model, where some developments have been recently made \cite{Queiruga4}, \cite{Gud1}, \cite{Gud2} . 
 
The last very interesting issue is related to a widely known fact that the nonlinear $O(3)$ $\sigma$-model in two dimensions (and also higher-dimensional target-space generalizations) is an integrable field theory with a zero-curvature formulation. It would be desirable to understand the fate of the integrability in twin supersymmetric extensions. Obviously, the theories remain integrable in their bosonic part but probably not necessarily when the fermions are included.

\section*{Acknowledgements}
AW was supported by NCN grant 2012/06/A/ST2/00396. We thank Wojtek Zakrzewski, Stefano Bolognesi and Christoph Adam for discussion.

\appendix

\end{document}